\documentclass[letterpaper,10pt,conference]{IEEEtran}

\usepackage{graphics} 
\usepackage{epsfig} 
\usepackage{mathptmx} 
\DeclareMathAlphabet{\mathcal}{OMS}{cmsy}{m}{n}
\usepackage{makecell}
\usepackage{tabularx}
\usepackage{adjustbox}
\usepackage{multirow}
\usepackage{array}
\newcolumntype{P}[1]{>{\centering\arraybackslash}p{#1}}
\usepackage{comment}
\usepackage{times} 
\usepackage{amsmath} 
\usepackage{amsfonts}
\usepackage{amssymb}  
\usepackage[version-1-compatibility]{siunitx}
\usepackage{bm}

\usepackage{enumitem}
\usepackage[hyphens]{url}
\usepackage[hidelinks]{hyperref}
\hypersetup{breaklinks=true}
\usepackage{cite}
\usepackage{soul}

\newtheorem{theorem}{Theorem}
\newtheorem{assumption}{Assumption}
\newtheorem{remark}{Remark}
\newtheorem{definition}{Definition}

\usepackage[T1]{fontenc}
\usepackage[utf8]{inputenc}
\usepackage[english]{babel}
\usepackage[dvipsnames]{xcolor}

\usepackage{graphicx}
\usepackage{xstring}
\usepackage{forloop}

\makeatletter
\newcommand*\bigcdot{\mathpalette\bigcdot@{1}}
\newcommand*\bigcdot@[2]{\mathbin{\vcenter{\hbox{\scalebox{#2}{$\m@th#1\bullet$}}}}}
\makeatother

\begin{document}

\newcommand\blfootnote[1]{%
  \begingroup
  \renewcommand\thefootnote{}\footnote{#1}%
  \addtocounter{footnote}{-1}%
  \endgroup
}

\title{\LARGE \bf
A Tethered Quadrotor UAV$-$Buoy System for Marine Locomotion}
\author{Ahmad~Kourani$^{1}$
        and Naseem Daher$^{2}$, ~\IEEEmembership{Member,~IEEE}
\thanks{$^{1}$Ahmad Kourani is with the Vision and Robotics Lab, Department of Mechanical Engineering, 
American University of Beirut, Beirut, Lebanon
        {\tt\small ahk42@mail.aub.edu}}%

\thanks{$^{2}$Naseem Daher is with the Vision and Robotics Lab, Department of Electrical \& Computer Engineering,
       {\tt\small nd38@aub.edu.lb}}%
       }%



       
\IEEEoverridecommandlockouts

\IEEEpubid{\makebox[\columnwidth]{\copyright{}2021 IEEE. Personal use of this material is permitted. \hfill} \hspace{\columnsep}\makebox[\columnwidth]{ }}      

\maketitle

\begin{abstract}
Unmanned aerial vehicles (UAVs) are finding their way into offshore applications. In this work, we postulate an original system that entails a marine locomotive quadrotor UAV that manipulates the velocity of a floating buoy by means of a cable. By leveraging the advantages of UAVs relative to high speed, maneuverability, ease of deployment, and wide field of vision, the proposed UAV$-$buoy system paves the way in front of a variety of novel applications. The dynamic model that couples the buoy, UAV, cable, and water environment is presented using the Euler-Lagrange method. A stable control system design is proposed to manipulate the forward-surge speed of the buoy under two constraints: maintaining the cable in a taut state, and keeping the buoy in contact with the water surface. 
Polar coordinates are used in the controller design process to attain correlated effects on the tracking performance, whereby each control channel independently affects one control parameter. This results in improved performance over traditional Cartesian-based velocity controllers, as demonstrated via numerical simulations in wave-free and wavy seas.
\end{abstract}



\section{Introduction} \label{introduction}

%
%
Unmanned aerial vehicles (UAVs) can be used for transmitting power \cite{Talke2018,Murison2018}, forces \cite{Kovac2019,SoRyeokOh2006}, and data \cite{Ozkan2019}. By coupling a UAV to a tether,
it can interact with its environment \cite{Nicotra2014}, which allows it to be more than a passive flying machine, and enables it to be incorporated in applications that require active manipulation of components in its vicinity.
%
%
\par
%
The limited power capacity and flight time of multi-rotor UAVs has resulted in a low engagement with the marine environment. Feasible applications are limited to information gathering such as visual reconnaissance,  identifying the locations of floating objects for retrieval missions \cite{Miskovic2014}, and drawing on-site maps and trajectories that allows other agents to perform rescue missions \cite{Ozkan2019}, among others. Limited physical interaction with the environment is also applicable to low-power tasks, including sensing jobs \cite{Tognon2019}, power-feeding the UAV through a cable  \cite{Talke2018,choi2014tethered}, and even landing assistance on a rocking ship \cite{SoRyeokOh2006}.
\par
Given the above limitations, unmanned surface vehicles (USVs) have been the choice for autonomous missions in marine environments. However, UAVs offer advantages relative to their bird's-eye view, high maneuverability, and ease of deployment, which allows them to outperform USVs in remote unstructured areas, and in tasks requiring agility and precision. Hence, having a tethered UAV$-$buoy system can enable new UAV-marine applications.
\par
For instance, remote oil slick thickness measurement has various limitations \cite{Fingas2018OilSlick}; for that an \textit{in situ} oil slick thickness sensor was designed in \cite{Saleh2019} to be deployed during marine oil spill events. The proposed sensor, fixed to a floating buoy, gets pulled by another vessel to skim the water surface while measuring the oil slick thickness. In such a configuration, the vessel ahead inevitably disturbs the oil layer resulting in reduced measurement accuracy.
Motivated by such marine applications, and benefiting from the maturity of the UAV technology,
this paper introduces the idea of employing a quadrotor UAV to manipulate a passive floating object by means of a cable.
The UAV$-$buoy system becomes even more appealing given that umbilical power line solutions, as in \cite{Talke2018}, naturally integrate into the proposed architecture since the cable can be used for both force and power transmission, where the UAV performs the locomotion task and the buoy carries the power banks for power efficiency considerations.
The proposed marine locomotive UAV problem generalizes the fixed-point tether described in \cite{Nicotra2014} and \cite{Tognon2017}, and the single-axis moving-frame tether in \cite{Nguyen2019}, to a double-axis moving-frame tether, with free motion in both the horizontal and vertical directions. Additionally, the nature of the problem poses additional constraints to the UAV$-$Buoy system, such as maintaining tension in the cable and contact with the water surface.
\par
%
%
The proposed system can be used in coordination with nearby ships and marine structures, thus increasing their maneuverability and decreasing their response time. It can also be beneficial nearshore and across other water surfaces such as rivers and waterfalls, which opens the way for a new set of solutions based on the interaction of UAVs and USVs.
\par
%
%
This paper provides several technical contributions. First, we propose a novel UAV-based solution for marine applications by formulating the problem of a marine locomotive UAV, which opens the door in front of further exploration into the interaction between UAVs and the marine environment.
Second, the system is defined in a sea/ocean environment with its dynamic model accounting for the presence of gravity waves and surface current, while considering appropriate constraints that can be tailored to fit the desired application.
Third, we design an application-specific control system that regulates the buoy's forward-surge velocity. The proposed controller outperforms traditional UAV-only velocity controllers in terms of tracking accuracy and performance, while reducing the system's energy consumption by maintaining a constant UAV altitude.
\par
%
%
The rest of this paper is structured as follows. Section \ref{sec_modeling} provides 
a description of the tethered UAV$-$buoy system model, followed by the designed control system in Section \ref{sec_controller_design}. Section \ref{sec_simulation} presents numerical simulation results that demonstrate the validity of the proposed system model and the effectiveness of the designed controller. Section \ref{sec_conclusion} concludes the paper and sets future areas of investigation. \par
%
%
%
\section{Tethered UAV-Buoy System Model} \label{sec_modeling} %
\subsection{Problem Formulation} \label{subsec_problem_statement}
\begin{figure}
\centerline{\includegraphics[width=3.40in]{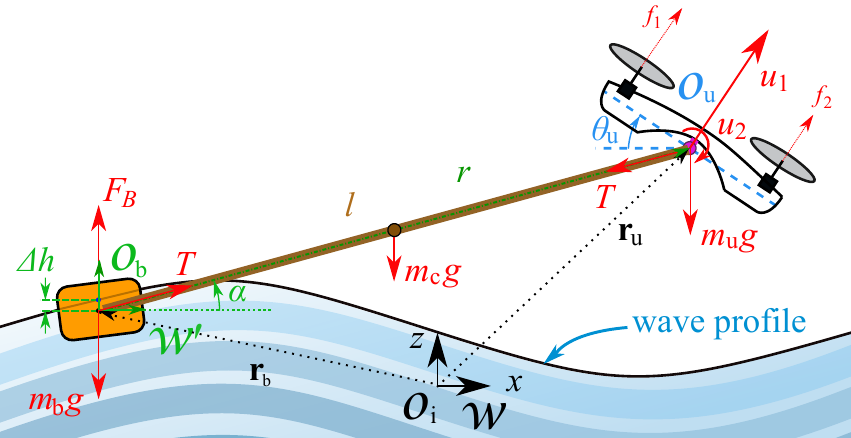}}
\caption{Planar model of a quadrotor UAV pulling a floating buoy through a tether.}
\label{fig_Buoy_UAV_Annotations}
\end{figure}

Consider the two-dimensional space in the water vertical plane where the problem is set up as shown in Fig.~\ref{fig_Buoy_UAV_Annotations}, and let $\mathcal{W}=\{x,y\}$ represent the inertial frame of reference whose origin, $\mathcal{O}_{\mathrm{i}}$, is at the local mean sea level horizontal line.
%
%
The floating buoy has a volume $ \curlyvee_{\mathrm{b}} \in \mathbb{R}_{>0}$, which is the set of positive-real numbers, mass $m_{\mathrm{b}} \in (0,\rho_{\mathrm{w}} \curlyvee_{\mathrm{b}})$, where $\rho_{\mathrm{w}}$ is the water density, and moment of inertia $J_{\mathrm{b}} \in \mathbb{R}_{>0}$. The quadrotor UAV has a mass $m_{\mathrm{u}}$ and moment of inertia $J_{\mathrm{u}}$. The cable, considered inextensible, has a length $l \in \mathbb{R}_{>0}$, mass $m_{\mathrm{c}}$, and moment of inertia $J_{\mathrm{c}}$.
The buoy is mechanically attached to the UAV by means of a cable, forming an angle $\alpha \in (0,\frac{\pi}{2})$ with the positive $x$-axis, which is defined as the elevation angle. 
Let $\mathbf{r}_{\mathrm{b}}=\{x_{\mathrm{b}},z_{\mathrm{b}}\} \in \mathbb{R}^2$ and $\mathbf{r}_{\mathrm{u}}=\{x_{\mathrm{u}},z_{\mathrm{u}}\} \in \mathbb{R}^2$ be the coordinates of the centers of mass of the buoy, ($\mathcal{O}_{\mathrm{b}}$), and the UAV, ($\mathcal{O}_{\mathrm{u}}$), in $\mathcal{W}$, respectively. For practical purposes, we set $V:=\dot{x}_{\mathrm{b}}$ to depict the buoy's horizontal velocity. 
Let $\mathcal{B}_{\mathrm{b}}$ and $\mathcal{B}_{\mathrm{u}}$ be the body-fixed reference frames of the buoy at $\mathcal{O}_{\mathrm{b}}$, and of the quadrotor at $\mathcal{O}_{\mathrm{u}}$, respectively.
Also let the orientation, measured clockwise, of $\mathcal{B}_{\mathrm{b}}$ and $\mathcal{B}_{\mathrm{u}}$ with respect to $\mathcal{W}$ be described by the angles $\theta_{\mathrm{b}}$ and $\theta_{\mathrm{u}} \in (-\pi,\pi]$, respectively. 
\par
Let  $\mathcal{W}'=\{r',\alpha'\}$ be a rectilinear moving polar frame fixed to $\mathcal{O}_{\mathrm{b}}$, in which we define the relative motion of the buoy and the UAV; this frame does not rotate, and it is parallel to the inertial frame $\mathcal{W}$.
The position of the UAV in $\mathcal{W}$ with respect to $\mathcal{W'}$ is defined as: $\mathbf{r}=\mathbf{r}_{\mathrm{u}}-\mathbf{r}_{\mathrm{b}} \in \mathbb{R}^2$, 
and we let its coordinates in $\mathcal{W}'$ be $\mathbf{r}'=\{r,\alpha\}$, such that $r = \|\mathbf{r}\|$ and $\alpha = \text{atan2}(z_{\mathrm{u}}-z_{\mathrm{b}},x_{\mathrm{u}}-x_{\mathrm{b}})$.
\par
The buoy, UAV, and cable are subject to gravitational acceleration, $g$, and the cable tension, $T \in \mathbb{R}_{\geq0}$ (the set of non-negative real numbers), affects both the buoy and UAV.
Moreover, the buoy is subjected to hydrostatic and hydrodynamic forces, and the UAV propulsion can be simplified in the planar case to only include the total thrust $u_1 \in \mathbb{R}_{\geq0}$ and a single torque $u_2 \in \mathbb{R}$ that induces pitch motion. 
\par
Let the inertia matrix of the buoy in $\mathcal{B}_{\mathrm{b}}$ be $\mathbf{M}'_{\mathrm{b}}=diag(m_{\mathrm{b}}+a_{11},m_{\mathrm{b}}+a_{33},J_{\mathrm{b}}+a_{55}) \in \mathbb{R}^{3\times3}$, where $a_{11}$, $a_{33}$, and $a_{55} \in \mathbb{R}_{\geq0}$ are the surge, heave, and pitch rate components of the generalized added mass matrix \cite{Fossen1995}.
The total damping term of the buoy in $\mathcal{B}_{\mathrm{b}}$ is expressed as:
\begin{equation} \label{eq_buoy_body_damping}
    \mathbf{D}'_{\mathrm{b}} = \mathbf{D}_{\mathrm{P}} + \mathbf{D}_{\mathrm{S}},
\end{equation}
\noindent where $\mathbf{D}_{\mathrm{P}}=diag(b_{11},b_{33},b_{55}) \in \mathbb{R}^{3\times3}$ is the radiation induced potential damping matrix with surge, heave, and pitch components, and $\mathbf{D}_{\mathrm{S}}=diag(D_{\mathrm{S},1},D_{\mathrm{S},2},D_{\mathrm{S},3}) \in \mathbb{R}^{3\times3}$ is the skin friction matrix.
We also define the inertia and damping matrices of the buoy in $\mathcal{W}$, respectively, as:
$\mathbf{M}_{\mathrm{b}} =  \mathbf{R}_{\theta_{\mathrm{b}}} \mathbf{M}'_{\mathrm{b}} \mathbf{R}_{\theta_{\mathrm{b}}}^{-1}$, and 
$\mathbf{D}_{\mathrm{b}} =  \mathbf{R}_{\theta_{\mathrm{b}}} \mathbf{D}'_{\mathrm{b}}\mathbf{R}_{\theta_{\mathrm{b}}}^{-1}$, 
where $\mathbf{R}_{\theta_{\mathrm{b}}}$ is the translational rotation matrix from $\mathcal{B}_{\mathrm{b}}$ to $\mathcal{W}$.
Finally, let $M_{\mathrm{b},ij}$ and $D_{\mathrm{b},ij}$ ($i,j=1,2,3$) be elements of $\mathbf{M}_{\mathrm{b}}$ and $\mathbf{D}_{\mathrm{b}}$, respectively. 
\subsection{Water Medium Model} \label{subsec_water_medium}
The problem is defined in a sea/ocean environment, where the aspects of interest are gravity waves and water surface current, the models of which are given next. \par
\subsubsection{Gravity Wave Model}
\begin{assumption} \label{assump_linear_waves_theory}
    The water depth is assumed to be much larger than the wavelength of gravity waves, thus linear wave theory is adopted \cite{faltinsen1990}. In addition, only waves with moderate amplitudes are considered as rough seas are not in the scope of this work, and wave direction is limited to be in the problem plane. 
\end{assumption} \par
Based on Assumption \ref{assump_linear_waves_theory}, the water elevation variation, $\zeta$, 
due to gravity waves is statistically described as \cite{faltinsen1990}:
\begin{equation} \label{eq_water_surface_elevation}
    \zeta(x,t) = \sum_{n}^{N} A_n \sin(d_n \omega_n t - k_n x + \sigma_n),
\end{equation}
\noindent where $A_n$ is the wave amplitude, $\omega_n$ is the circular frequency, $k_n \in \mathbb{R}_{\geq 0}$ is the wave number, $d_n \in \{-1,1\}$ is the wave direction coefficient, and $\sigma_n \in (-\pi,\pi]$ is the random phase angle of wave component number $n \in S_n$ where $S_n = \{1 \leq n \leq N \, | \, N  \in \mathbb{N}\}$. Furthermore, based on Assumption \ref{assump_linear_waves_theory}, the wave number in deep water is given by the dispersion relation as $k_n = \omega_n^2/g$.
The horizontal and vertical fluid particles' wave-induced velocities can be prescribed as:
\begin{equation} \label{eq_wave_vel}
    \begin{split}
        v^w_{x}(x,z,t) & = \sum_{n}^{N} d_n \omega_n A_n e^{k_n z}   \sin(d_n \omega_n t - k_n x + \sigma_n),\\
        v^w_{z}(x,z,t) & = \sum_{n}^{N} d_n \omega_n A_n e^{k_n z}   \cos(d_n \omega_n t - k_n x + \sigma_n).\\
    \end{split}
\end{equation}
\par
\subsubsection{ Water Current}
The horizontal ($x$-direction) water surface current is given as:
\begin{equation} \label{eq_current}
    U_{\mathrm{cr}} = U_{\mathrm{l}} + U_{\mathrm{s}}, 
\end{equation}
\noindent where $U_{\mathrm{l}} \in \mathbb{R}$ is the lumped sum of different water current components, and $U_{\mathrm{s}} \in \mathbb{R}$ is the component generated from Stokes drift \cite{Fossen1995}.
\par
%
%
\subsection{System Constraints} \label{subsec_system_constraints}
To fully define the marine locomotive UAV problem as a coupled UAV$-$buoy system, certain constraints are required and are presented hereafter.
%
\subsubsection{Taut-Cable Constraint} \label{susubbsec_positive_cable_tension}
\begin{assumption} \label{assump_cable}
    The cable is inextensible and it is attached to the buoy's center of mass at one end, and to the UAV's center of mass at the other.
\end{assumption}
\begin{definition} \label{def_taut_cable}
    Based on Assumption~\ref{assump_cable}, the cable remains taut, i.e. maintains tension, at time $t$ if $r(t) = l$. The taut-cable condition is expressed as:
\begin{equation} \label{eq_taut_cable_condition} 
    T > 0.
\end{equation}
\end{definition}
\par
With Assumption \ref{assump_cable} and the taut-cable condition in (\ref{eq_taut_cable_condition}), we have $r=l$.
\par
%
\subsubsection{No Buoy-Hanging Constraint} \label{subsubsec_no_buoy_hanging_constraint}
The UAV must not lift the buoy into the air by means of the cable tension alone. This constraint can be forced by limiting the allowed cable tension by the following inequality: 
\begin{equation} \label{eq_no_buoy_hanging_constraint} 
    T < (m_{\mathrm{b}}+m_{\mathrm{c}}) g / \sin{\alpha}.
\end{equation}
%
\subsubsection{No `Fly Over' Constraint} \label{subsubsec_no_fly_over_constraint}
The buoy must maintain contact with the water surface at all times, that is, the UAV should not cause the buoy to jump over the encountered waves \cite{Fridsma1969PlaningBoats}. This constraint is described as:
\begin{equation} \label{eq_water_contact_condition} 
    \curlyvee_{\text{im}} > 0,
\end{equation}
\noindent which guarantees keeping the buoy partially immersed. 
\par
%
\subsection{Euler-Lagrange Formulation}
The formulation of the UAV$-$buoy tethered system can be obtained by the Euler-Lagrange formulation.
%
Assuming that the buoy dynamics are stable and damped, and under constraints (\ref{eq_taut_cable_condition}) and (\ref{eq_water_contact_condition}), the dynamic model equations are given by \cite{Fossen1995,Kourani2018_Quadrotor_ARC}:
\begin{subequations}
\label{eq_dynamic_model_expended}
\begin{IEEEeqnarray}{rCl}  
    \IEEEeqnarraymulticol{3}{l}{
        (M_{\mathrm{b},11}+m_{\mathrm{u}}+m_{\mathrm{c}})\ddot{x}_{\mathrm{b}} + M_{\mathrm{b},12}\ddot{z}_{\mathrm{b}}  +  D_{\mathrm{b},11} V_r + D_{\mathrm{b},12} \tilde{\dot{z}}_{\mathrm{b}}
        }
        \label{eq_dynamic_model_expended_a}\\ \quad
    - M_a (c_{\alpha}\dot{\alpha}^2 + s_{\alpha}\ddot{\alpha}) & = & u_1 s_{\theta_{\mathrm{u}}}, \nonumber\\* 
    \IEEEeqnarraymulticol{3}{l}{
        (M_{\mathrm{b},22}+m_{\mathrm{u}}+m_{\mathrm{c}})\ddot{z}_{\mathrm{b}} + M_{\mathrm{b},21}\ddot{x}_{\mathrm{b}} 
        }
        \label{eq_dynamic_model_expended_b}\\ \quad
        - M_a(s_{\alpha}\dot{\alpha}^2 - c_{\alpha}\ddot{\alpha}) + D_{\mathrm{b},22} \tilde{\dot{z}}_{\mathrm{b}} & + & D_{\mathrm{b},21} V_r \nonumber\\
     + (m_{\mathrm{b}}+m_{\mathrm{u}}+m_{\mathrm{c}})g & = & u_1 c_{\theta_{\mathrm{u}}} + (\rho_{\mathrm{w}} \curlyvee_{\text{im}}) g,  \nonumber\\
    %
    J_a \ddot{\alpha} + M_a( -s_{\alpha}\ddot{x}_{\mathrm{b}} + c_{\alpha}\ddot{z}_{\mathrm{b}} + g c_{\alpha}) 
    & = & u_1 l c_{\alpha+\theta_{\mathrm{u}}}, \label{eq_dynamic_model_expended_c}\\
    J_{\mathrm{u}} \ddot{\theta}_{\mathrm{u}} & = & u_2, \label{eq_dynamic_model_expended_d}
\end{IEEEeqnarray}  
\end{subequations}
\noindent where $s_{\bigcdot}$ and $c_{\bigcdot}$ are the sine and cosine functions, respectively; 
$V_r = V-U_{\mathrm{cr}}-v_x^w$ is the buoy$-$water relative surge velocity; $\tilde{\dot{z}}_{\mathrm{b}}=\dot{z}_{\mathrm{b}}-v_z^w$; $M_a=m_{\mathrm{u}} l +m_{\mathrm{c}} \frac{l}{2}$; and $J_a = m_{\mathrm{u}} l^2 +m_{\mathrm{c}} \frac{l^2}{3}$.
The buoy's pitch angle is determined by differentiating (\ref{eq_water_surface_elevation}) with respect to $x_{\mathrm{b}}$:
\begin{equation} \label{eq_buoy_tilt_angle}
        \theta_{\mathrm{b}} = 
        \mathrm{atan}\Big( \sum_{n}^{N} A_n k_n \cos(d_n \omega_n t - k_n x_{\mathrm{b}} + \sigma_n)
        \Big).    
\end{equation}
\par
Using a polar coordinates notation, the UAV's equations of motion in $\mathcal{W}'$ are expressed as:
\begin{equation} \label{eq_UAV_polar_frame} 
    \begin{split}
        - m_{\mathrm{u}} l \dot{\alpha}^2 & = m_{\mathrm{u}} ( - \ddot{x}_{\mathrm{b}} c_{\alpha} - \ddot{z}_{\mathrm{b}} s_{\alpha}) - m_{\mathrm{u}} g s_{\alpha} + u_1 s_{\alpha + \theta_{\mathrm{u}}} - T, \\
            m_{\mathrm{u}} l^2 \ddot{\alpha} & = m_{\mathrm{u}} l ( \ddot{x}_{\mathrm{b}} s_{\alpha} - \ddot{z}_{\mathrm{b}} c_{\alpha} ) - m_{\mathrm{u}} g l c_{\alpha} + l u_1 c_{\alpha + \theta_{\mathrm{u}}},
    \end{split}
\end{equation}
\noindent where the cable tension can be derived from the buoy's equations of motion as:
\begin{equation} \label{eq_buoy_cable_tension_X} 
        T = \Big(M_{\mathrm{b},11} \ddot{x}_{\mathrm{b}} + M_{\mathrm{b},12} \ddot{z}_{\mathrm{b}}
        + D_{\mathrm{b},11} V_r + D_{\mathrm{b},12} \tilde{\dot{z}}_{\mathrm{b}} 
        \Big) / c_{\alpha}.
\end{equation}    
%
%
\section{Control System Design} \label{sec_controller_design} 
Consider the system in (\ref{eq_dynamic_model_expended}) and (\ref{eq_UAV_polar_frame}), the control problem is defined as manipulating two variables: the surge speed of the buoy, $V$, and the cable angle, $\alpha$, while ensuring stability of the UAV$-$buoy system dynamics and respecting constraints (\ref{eq_taut_cable_condition}) - (\ref{eq_water_contact_condition}).
\par
%
\subsection{Control System Design} 
\begin{figure}
\centerline{\includegraphics[width=3.4in]{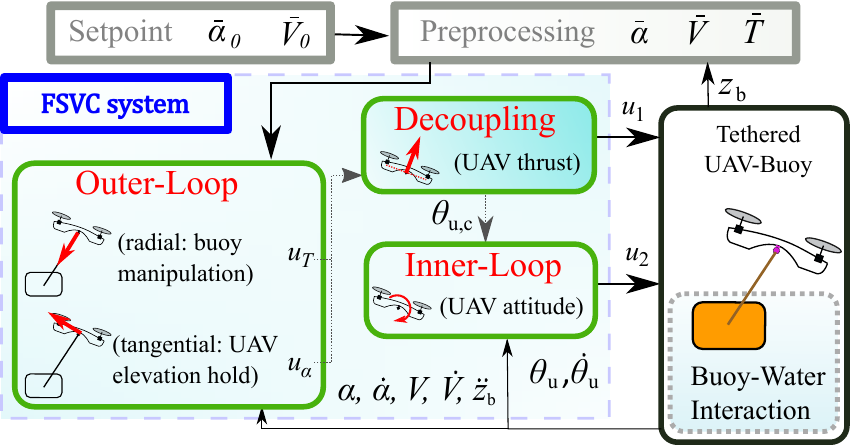}}
\caption{The forward-surge velocity control (FSVC) system architecture.} 
\label{fig_Controller_Diagram}
\end{figure}

The control system design consists of an outer-loop controller that regulates the cable tension and elevation angle, labelled as forward-surge velocity control (FSVC) system, and an inner-loop controller that controls the UAV's pitch angle. 
The FSVC system design is based on polar coordinates, and its architecture is presented in Fig.~\ref{fig_Controller_Diagram}.
\par
At this point, it is noteworthy to mention that the use of a nominal Cartesian-based proportional-integral-derivative (PID) outer-loop controller resulted in a marginal performance, as later shown in Section \ref{sec_simulation}.
%
\par
\subsubsection{Reference Signals and Velocity Setpoint}
Let $\bar{\alpha}_0$ be the nominal elevation angle. In order to maintain a level flight,
a corrected elevation angle, $\bar{\alpha}$, can be computed as:
\begin{equation} \label{eq_alpha_desired}
    \bar{\alpha}=\text{asin}\big((\bar{z}_{\mathrm{u}}-z_{\mathrm{b}})/l \big),
\end{equation}
\noindent where $\bar{z}_{\mathrm{u}}$ is the reference UAV's height command.
Furthermore, let $\bar{V}_0$ be the velocity setpoint, which is smoothed by a second-order low-pass filter to become:
\begin{equation} \label{eq_V_desired}
    \bar{V} = \frac{1}{(\tau_f s + 1)^2} \bar{V}_0,
\end{equation}
\noindent where $\tau_f \in \mathbb{R}_{>0}$ is the filter's time constant. A smoothed  velocity profile respects the system dynamics in terms of buoy$-$water friction and the UAV's maximum thrust, and consequently prevents constraints violation.
\par
\subsubsection{Outer-Loop Controller}
Consider the UAV dynamics in $\mathcal{W}'$ while following the polar coordinates notation presented in (\ref{eq_UAV_polar_frame}), and let the buoy's velocity and cable's elevation angle errors be respectively defined as:
            $e_V = V-\bar{V}$ and $e_{\alpha} = \alpha-\bar{\alpha}$.
%
The buoy velocity and cable elevation angle models are expressed as:
\begin{equation} \label{eq_velocity_elevation_model}
    \begin{split}
        \dot{V}  & =   H_T  + (u_T - T)/ (m_{\mathrm{u}}  c_{\alpha}), \\
        \ddot{\alpha} & = H_{\alpha}  + u_{\alpha} / (m_{\mathrm{u}} l),
    \end{split}
\end{equation}
\noindent where
    $H_T = ( l \dot{\alpha}^2 - \ddot{z}_{\mathrm{b}} s_{\alpha} - g s_{\alpha}) / c_{\alpha}$,
    $H_{\alpha} = (\dot{V} s_{\alpha} - \ddot{z}_{\mathrm{b}} c_{\alpha} - g c_{\alpha}) / l$,
    $u_T = u_1 s_{\alpha + \theta_{\mathrm{u}}}$, and
    $u_{\alpha} = u_1 c_{\alpha + \theta_{\mathrm{u}}}$.
\noindent Note that $H_T$ and $H_{\alpha}$ represent the nonlinear and gravitational terms in each channel. However, the cable tension cannot be exactly known, and the system may be subjected to unknown external disturbances like wind gusts, gravity waves, and water currents. Hence, (\ref{eq_velocity_elevation_model}) is rewritten in the following form:
    \begin{equation} \label{eq_velocity_elevation_model_disturbed}
        \begin{split}
            \dot{V}  & =   H_T  + (u_T - \hat{T})/ (m_{\mathrm{u}}  c_{\alpha}) + \delta_V, \\
            \ddot{\alpha} & = H_{\alpha}  + u_{\alpha} / (m_{\mathrm{u}} l) + \delta_{\alpha},
        \end{split}
    \end{equation}
\noindent where $\hat{T}$ is the cable tension estimate formulated from (\ref{eq_buoy_cable_tension_X}); $\delta_V$ and $\delta_{\alpha}$ are the lumped system disturbances and modeling errors across each channel, and $\hat{\delta}_V$ and $\hat{\delta}_{\alpha}$ denote their estimates, respectively.
\begin{assumption}\label{assump_bounded_dist}
  The modeling errors and external disturbances and their derivatives are bounded, 
  and the lumped errors $\delta_{\alpha}$ and $\delta_{V}$ are constant or slowly varying during a finite time interval, that is: $\lim_{t_1<t<t_2} \dot{\delta}_{\alpha}, \dot{\delta}_{V} \approx 0$. 
\end{assumption}
\par
Let $\theta'_{\mathrm{u,c}}$ be the desired UAV pitch angle to be generated by the outer-loop controller along with the total thrust command, $u_1$, which are calculated as:
\begin{equation} \label{eq_outer_loop_u1_thetac}
       u_{1} = \sqrt{u_{\alpha}^2+u_{T}^2}, \quad
        \theta'_{\mathrm{u,c}} = \frac{\pi}{2}-\alpha - \text{arctan}(u_{\alpha},u_T),
\end{equation}
\noindent where the radial and tangential thrust components, $u_T$ and $u_{\alpha}$, respectively, are defined as \cite{Kourani2021_PID_Like_Backstepping}:
\begin{equation} \label{eq_outer_loop_u_alpha_T}
    \begin{split}
        u_T & = \hat{T} + m_{\mathrm{u}} c_{\alpha} \big( -( l \dot{\alpha}^2 - \ddot{z}_{\mathrm{b}} s_{\alpha} - g s_{\alpha}) / c_{\alpha}  + \dot{\bar{V}} \\
        & \quad - k_{PV} e_{V} - k_{IV} e_{V}^I \big), \qquad \qquad \;\;\;
        \dot{e}_{V}^I = e_V,\\
        u_{\alpha} & = m_{\mathrm{u}} l \big( - (\dot{V} s_{\alpha} - \ddot{z}_{\mathrm{b}} c_{\alpha} - g c_{\alpha}) / l  + \ddot{\bar{\alpha}} \\
        & \quad -  k_{P\alpha}e_{\alpha} - k_{D\alpha} \dot{e}_{\alpha} - k_{I \alpha} e_{\alpha}^I \big), \quad
        \dot{e}_{\alpha}^I = e_{\alpha} + k_{\alpha1}^{-1} \dot{e}_{\alpha},
    \end{split}
\end{equation}
\noindent where $k_{P \alpha}$, $k_{D \alpha}$, $k_{I\alpha}$, $k_{PV}$, and $k_{IV}$ are tuning gains that are defined next.
\begin{theorem} \label{theorem_outer_loop_controller}
    Consider the tethered UAV$-$buoy system model described in (\ref{eq_dynamic_model_expended}), and the buoy velocity and elevation angle dynamics expressed in (\ref{eq_velocity_elevation_model_disturbed}). Suppose that Assumption \ref{assump_bounded_dist} holds true; the control law in (\ref{eq_outer_loop_u1_thetac}) and (\ref{eq_outer_loop_u_alpha_T}) generates the total thrust, $u_1$, and the UAV's desired pitch angle, $\theta_{\mathrm{u,c}}'$, which can stabilize the system and reduce the tracking error asymptotically to zero for a set of gains $k_{\alpha1}$, $k_{\alpha2}$, $k_{V}$, $\gamma_{\alpha}$, and $\gamma_V \in \mathbb{R}_{>0}$,
    such that $k_{P \alpha} = 1+k_{\alpha1} k_{\alpha2}$, $k_{D \alpha} = k_{\alpha1} + k_{\alpha2}$, $k_{I\alpha} = \gamma_{\alpha} k_{\alpha1}$, $k_{PV}=k_{V}$, and $k_{IV} = \gamma_{V}$. 
\end{theorem}
\begin{IEEEproof}
    The backstepping control design algorithm is employed with the following two Lyapunov functions: $\mathcal{V}_V=\frac{1}{2} e_V^2+\frac{1}{2 \gamma_V} \tilde{\delta}_V^2$ where $\tilde{\delta}_V=\hat{\delta}_V - \delta_V$,
    and $\mathcal{V}_{\alpha1}=\frac{1}{2} e_{\alpha}^2$. The derivatives of the Lyapunov functions are expressed as: 
    \begin{equation} \label{eq_VValpha1_dot}
            \dot{\mathcal{V}}_V = e_V\dot{e}_V + \tilde{\delta}_V \dot{\hat{\delta}}_V / \gamma_V, \quad
            \dot{\mathcal{V}}_{\alpha1} = e_{\alpha}\dot{e}_{\alpha}.
    \end{equation}
    Since $\dot{e}_{\alpha1}$ does not explicitly include a control input, we continue the control design process for a second step in the elevation angle channel. The virtual control input to stabilize $e_{\alpha}$ is defined as: $\Omega_{\alpha} = \dot{\bar{\alpha}} - k_{\alpha1} e_{\alpha}$.
    Next, we define the virtual elevation angular rate error as:  $e_{\Omega_\alpha} = \dot{\alpha} - \Omega_{\alpha}$.
    \par
    By defining another Lyapunov function for the elevation angle channel:
        $ \mathcal{V}_{\alpha2} = \frac{1}{2} e_{\alpha}^2 + \frac{1}{2}e_{\Omega_{\alpha}}^2 + \frac{1}{2 \gamma_{\alpha}} \tilde{\delta}_{\alpha}^2 $,
    %
    where $\tilde{\delta}_{\alpha}=\hat{\delta}_{\alpha} - \delta_{\alpha}$,
    then differentiating and combining it with (\ref{eq_VValpha1_dot}), we get:
    \begin{align*}
        \dot{\mathcal{V}}_V & = e_V \big(H_T  + (u_T - \hat{T})/ (m_{\mathrm{u}}  c_{\alpha}) + \delta_V - \dot{\bar{V}} \big) + \tilde{\delta}_V \dot{\hat{\delta}}_V / \gamma_V, \\
        \dot{\mathcal{V}}_{\alpha2} & = e_{\alpha} \dot{e}_{\alpha} + e_{\Omega_{\alpha}} \dot{e}_{\Omega_{\alpha}} + \tilde{\delta}_{\alpha} \dot{\hat{\delta}}_{\alpha} / \gamma_{\alpha} \\ 
        & = e_{\alpha}(e_{\Omega_{\alpha}} - k_{\alpha1}e_{\alpha}) + e_{\Omega_{\alpha}} (H_{\alpha}  + u_{\alpha} / (m_{\mathrm{u}} l) + \delta_{\alpha} - \dot{\Omega}_{\alpha} ) \\
        & \quad + \tilde{\delta}_{\alpha} \dot{\hat{\delta}}_{\alpha} / \gamma_{\alpha}.
    \end{align*}    
    Next, we choose the control inputs and the lumped modeling and disturbances errors' update rates, such that $\dot{\mathcal{V}}_V$ and $\dot{\mathcal{V}}_{\alpha2}$ become negative semi-definite:
    \begin{equation} \label{eq_U_T_alpha_lyapunov}
        \begin{split}
            u_T  & = \hat{T} + m_{\mathrm{u}} c_{\alpha} (-H_T - \hat{\delta}_V + \dot{\bar{V}} - k_V e_V), \qquad \;\;\;
            \dot{\hat{\delta}}_{V} = \gamma_V e_V, \\
            u_{\alpha}  & = m_{\mathrm{u}} l \big( -H_{\alpha} -\hat{\delta}_{\alpha} + \dot{\Omega}_{\alpha} - e_{\alpha} - k_{\alpha2} e_{\Omega_{\alpha}} \big), \quad  
            \dot{\hat{\delta}}_{\alpha} = \gamma_{\alpha} e_{\Omega_{\alpha}},
        \end{split}
    \end{equation}
    which yields $\dot{\mathcal{V}}_V = -k_V e_V^2$ and $\dot{\mathcal{V}}_{\alpha} = -k_{\alpha1}e_{\alpha}^2 -k_{\alpha2}e_{\Omega_{\alpha}}^2$. 
    Thus, asymptotic convergence of $\mathcal{V}_V$ and $\mathcal{V}_{\alpha}$ to zero can be obtained via Barbalat's lemma under Assumption~\ref{assump_bounded_dist}.
    Finally, by substituting $\dot{\Omega}_{\alpha}$ and $e_{\Omega_{\alpha}}$ in (\ref{eq_U_T_alpha_lyapunov}), and setting $e_V^I := \hat{\delta}_V/\gamma_V$ and $e_{\alpha}^I := \hat{\delta}_{\alpha} / (\gamma_{\alpha} k_{\alpha1})$, the PID-like control law in (\ref{eq_outer_loop_u_alpha_T}) is obtained.
\end{IEEEproof}
\par
\subsubsection{Inner-Loop Controller}
Let $\theta_{\mathrm{u,c}}=\theta_{\mathrm{u,m}} \, \text{tanh}\big(\theta'_{\mathrm{u,c}}/\bar{\theta}_{\mathrm{u,c}}\big)$
be a smooth and bounded version of $\theta'_{\mathrm{u,c}}$, with $ \theta_{\mathrm{u,m}} \in  (0,\frac{\pi}{2})$ being the absolute upper limit of the UAV's attitude angle, and let $e_{\theta_{\mathrm{u}}} = \theta_{\mathrm{u}} - \theta_{\mathrm{u,c}}$ be the UAV's attitude error.
%
    Considering the quadrotor UAV attitude system described in (\ref{eq_dynamic_model_expended_d}), the following control law can stabilize its dynamics based on Theorem~\ref{theorem_outer_loop_controller}:
    \begin{equation} \label{eq_U_attitue}
       u_2 = J_{\mathrm{u}}(-k_{P \theta} e_{\theta_{\mathrm{u}}} -k_{D \theta} \dot{e}_{\theta_{\mathrm{u}}} + \ddot{\theta}_{\mathrm{u,c}} ),
    \end{equation}
    \noindent where $k_{P \theta} = 1+k_{\theta1} k_{\theta2}$ and $k_{D \theta} = k_{\theta1} + k_{\theta2}$, with $k_{\theta1}$ and $k_{\theta2} \in \mathbb{R}_{>0}$ are tuning control parameters. 
    Furthermore, the tracking error is guaranteed to converge to zero in finite time via Theorem~\ref{theorem_outer_loop_controller}.
\par 
%

\section{Simulations} \label{sec_simulation}
%
\subsection{Simulation Settings}
 
Simulations are performed in the MATLAB Simulink\,\textsuperscript{\tiny\textregistered} environment to validate the proposed UAV$-$buoy system model and the designed FSVC system. This is made publicly available\footnote{\texttt{github.com/AUBVRL/Primitive-ML-UAV}}.
The buoy shape is simplified as a homogeneous cuboid with the dimensions and parameters listed in Table~\ref{tab_system_param}. The quadrotor UAV motor dynamics are neglected.
The mass of the buoy is chosen such that the buoy is one quarter immersed under no external loads based on the balance between gravitational and buoyancy forces, that is $m_{\mathrm{b}} := \rho_{\mathrm{w}} \curlyvee_{\mathrm{b}}/4$.
%
\begin{table}
    \caption{Tethered UAV$-$buoy system model parameters}
    \begin{center}
    \begin{tabular}{ p{1.15cm} p{1.2cm} p{0.8cm} |p{1.15cm} p{0.9cm} p{0.8cm} } 
        \hline 
        Parameter & Value & Unit & Parameter & Value & Unit \\
        \Xhline{2\arrayrulewidth}
        \hline
        $l_{\mathrm{b}}$  &  0.8  &  $\SI[unitsep=medium]{}{m}$   & $m_{\mathrm{u}}$  &  1.8  &   $\SI[unitsep=medium]{}{kg}$  \\
        $h_{\mathrm{b}}$  &  0.25 &  $\SI[unitsep=medium]{}{m}$    & $J_{\mathrm{u}}$  &  0.03  &  $\SI[unitsep=medium]{}{kg.m^2}$ \\
        $m_{\mathrm{b}}$  & 12.5   & $\SI[unitsep=medium]{}{kg}$    & $\theta_{\mathrm{u,m}}$  &  $\pi/4$  &  $\SI[unitsep=medium]{}{\radian}$ \\
        $a_{11}$  &  $0.625$  & $\SI[unitsep=medium]{}{kg}$    & $l$  &  7 & $\SI[unitsep=medium]{}{m}$\\
        $a_{33}$  &  $12.5$  & $\SI[unitsep=medium]{}{kg}$ &  $m_{\mathrm{c}}$  &  0.5  &   $\SI[unitsep=medium]{}{kg}$ \\
        $b_{11}$  &  0  & $\SI[unitsep=medium]{}{N.s /m}$ & $\tau_f$  &  2  &  \SI[unitsep=medium]{}{-}\\
        $b_{33}$  & 27.5  & $\SI[unitsep=medium]{}{N.s /m}$  & $g$  & 9.81 & $\SI[unitsep=medium]{}{m/s^2}$ \\      
        $ C_{\mathrm{S},1}$  &  5$\times$10$^{-3}$  & \SI[unitsep=medium]{}{-} & $\rho_{\mathrm{w}}$  &  1000  & $\SI[unitsep=medium]{}{kg/m^3}$ \\
        $ C_{\mathrm{S},2}$ & 9$\times$10$^{-3}$ & $\SI[unitsep=medium]{}{-}$ & \SI[unitsep=medium]{}{-}  &  \SI[unitsep=medium]{}{-}  & \SI[unitsep=medium]{}{-} \\ 
        \hline
            \end{tabular}
    \label{tab_system_param}
    \end{center}
\end{table}
%
The buoy's immersed volume is bounded such that $\curlyvee_{\mathrm{im}} \in [0,\curlyvee_{\mathrm{b}}]$.
The skin friction components are calculated as $D_{\mathrm{S},1} = C_{\mathrm{S},1} A_{\mathrm{wt}} \frac{1}{2} \rho_{\mathrm{w}} |V_r|$ and  $D_{\mathrm{S},2} = C_{\mathrm{S},2} A_{\mathrm{wt}} \frac{1}{2} \rho_{\mathrm{w}} |\tilde{\dot{\mathrm{z}}}_b|$, where $A_{\mathrm{wt}} \in [0,4 l_{\mathrm{b}} h_{\mathrm{b}}]$ is the buoy's wetted area.
\par
The proposed FSVC system is compared to a baseline PID outer-loop controller for benchmarking purposes. The nominal PID design consists of a velocity ($\dot{x}$) controller and an elevation ($z$) controller, and it was fine-tuned to minimize the tracking errors.
Let $\bm{k}_{s}=[k_{Ps},k_{Is},k_{Ds}]$ be a set of controller gains, where subscript $(\;)_{s}$ refers to a single controller channel.
The tuned gains are: $\bm{k}_{\dot{x}}=[7,1.2,5]$ for velocity, $\bm{k}_{z}=[3,1,2]$ for elevation, $\bm{k}_{\alpha}=[7,2,7]$ for tangential thrust, $\bm{k}_{T}=[60,9.6,0]$ for radial thrust, and $\bm{k}_{\theta}=[7.8,0,5.4]$ for pitch control.
The feedback signals are assumed to be known for proof-of-concept purposes.
\subsection{Simulation Scenario}
\begin{figure}
\centerline{\includegraphics[width=3.4in]{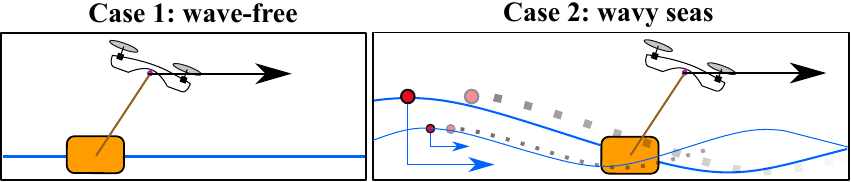}}
\caption{Depiction of the two simulated scenarios: wave-free (left) and wavy following seas (right).}
\label{fig_simulation_scenarios}
\end{figure}

Two cases are considered:
\begin{itemize}
    \item $C1$, wave-free: $N=1$ and $A_1=\SI[unitsep=medium]{0}{m}$.
    \item $C2$, moderate following seas ($N=2$). 
    The first wave component characteristics are: $A_1=\SI[unitsep=medium]{0.75}{m}$, $d_1=1$, $\omega_1=(2 \pi)/\SI[unitsep=medium]{5.7}{\rad}$, and $\sigma_1=0$; the second wave component characteristics are: $A_2=\SI[unitsep=medium]{0.135}{m}$, $d_2=1$, $\omega_2=(2 \pi)/\SI[unitsep=medium]{3}{\rad}$, and $\sigma_2=\pi$.

\end{itemize}
\noindent The two environments can be visualized in Fig.~\ref{fig_simulation_scenarios}.
In both cases, the buoy is requested to accelerate to reach an inertial velocity of $\bar{V}=\SI[unitsep=medium]{5}{m/s}$ and then to decelerate to $\SI[unitsep=medium]{3}{m/s}$. Note that the velocity is lower bounded by the \textit{Taut-Cable} constraint (\ref{eq_taut_cable_condition}), and is upper bounded by the \textit{No Buoy-Hanging} (\ref{eq_no_buoy_hanging_constraint}) and the \textit{No Fly Over} (\ref{eq_water_contact_condition}) constraints.
The mean elevation angle is set to $\bar{\alpha}_0 = \SI[unitsep=medium]{45}{\degree}$, which implies a desired reference mean sea level altitude of $\bar{z}_{\mathrm{u}}= \frac{h_{\mathrm{b}}}{4} + l s_{\alpha}=\SI[unitsep=medium]{5.0}{m}$.
The system's velocity is initiated to be equivalent to the zero-time water velocity as calculated from (\ref{eq_wave_vel}) and (\ref{eq_current}).
\par
%
%
\subsection{Simulation Results}
\begin{figure} 
\centerline{\includegraphics[width=3.4in]{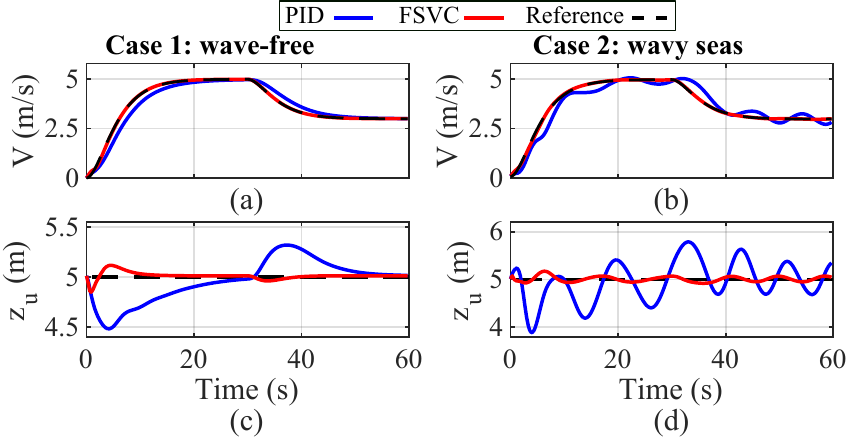}}
\caption{Buoy velocity (top row) and UAV's elevation variation (bottom row) during the buoy's locomotion task for simulation cases $C1$ (left) and $C2$ (right). The performance of the FSVC system (red) is compared against that of a baseline PID controller (blue). Reference trajectory is in black.}
\label{fig_buoy_vel_uav_elev}
\end{figure}

The state outputs of the system, namely the buoy's surge velocity and the UAV's elevation are shown in Fig.~\ref{fig_buoy_vel_uav_elev}.
%
The quadrotor UAV equipped with the proposed FSVC system is able to pull the buoy at the desired velocities in both wave-free and wavy seas, with acceptable transient and steady-state performance while reducing fluctuations in velocity and elevation.
On the other hand, a fine-tuned baseline PID controller 
has a marginal performance in maintaining a steady velocity and level flight, especially in wavy-seas conditions, resulting in large velocity and elevation fluctuations that reach up to $\SI[unitsep=medium]{0.4}{m/s}$ and $\SI[unitsep=medium]{1.2}{m}$, respectively. 
\par
A quantitative assessment of the two controllers is given in Table~\ref{tab_tracking_error_energy}. The last column shows the energy consumed by each quadrotor UAV based on \cite{Morbidi2016MinimumEnergy, [Christoph_CDC]}. The FSVC system achieves more than $90\%$ average tracking error reduction as compared to the PID controller, with an increase of less than $2\%$ in energy consumption, which is relatively minimal given the attained motion control accuracy.
%
It is also evident from Fig.~\ref{fig_buoy_vel_uav_elev}(b), that the PID controller does note cope well with water velocity fluctuations induced by the sea waves due to its smaller corrections, which translate to slightly lower energy consumption. However, this should not conceal the fact that the large vertical motion fluctuation of the UAV, shown in Figs.~\ref{fig_buoy_vel_uav_elev}(c)-(d),  deteriorates the power efficiency of the PID controller since the resultant energy gets dissipated.
\par
\begin{figure} 
\centerline{\includegraphics[width=3.4in]{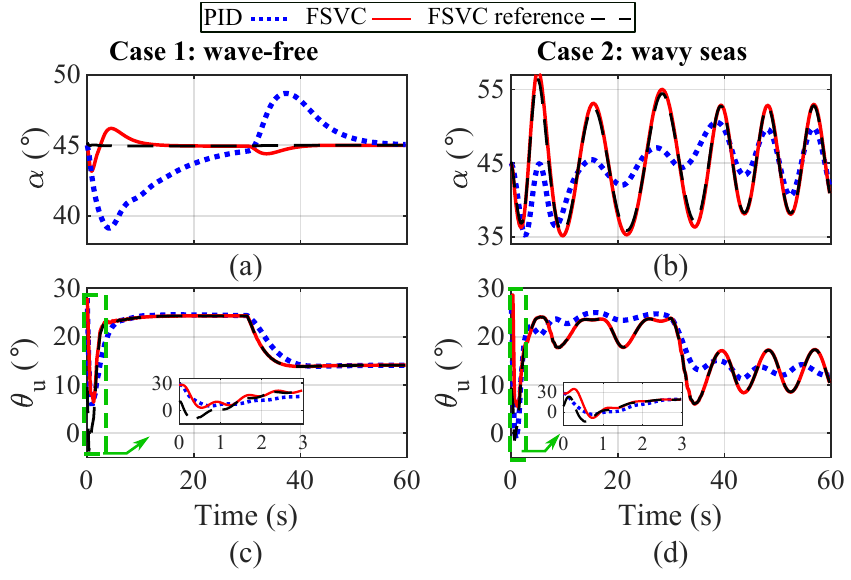}} 
\caption{Main system states, elevation angle $\alpha$ and pitch angle $\theta_{\mathrm{u}}$, during simulation with the designed FSVC system (red) and a baseline PID controller (blue).} 
\label{fig_alpha_theta}
\end{figure}

To avoid unnecessary vertical motion of the UAV, the FSVC system adjusts the elevation angle when the buoy elevation changes, which is evident in Fig.~\ref{fig_alpha_theta}(b).
It is also observed in Fig.~\ref{fig_alpha_theta} that the elevation ($\alpha$) and pitch ($\theta_{\mathrm{u}}$) angles are smooth and stable for both controllers, albeit the FSVC system results in zero tracking error, unlike the baseline PID controller.
Also, note that the oscillatory behavior of the elevation angle in $C2$, seen in Fig.~\ref{fig_alpha_theta}(b), is the natural response of the FSVC system under long waves excitation, where the buoy exhibits a contour-following behavior.

\begin{remark} \label{rem_correlated_effect}
    The formulation of the proposed controller in a polar coordinate system yields a correlated control performance, in which each control channel independently affects one control parameter ($\alpha$ or $V$). On the other hand, a Cartesian-based controller with $x$- and $z$-control channels produces dual effect on each control parameter, leading to a degraded performance. 
\end{remark}
\par
%
%
\begin{figure}
\centerline{\includegraphics[width=3.4in]{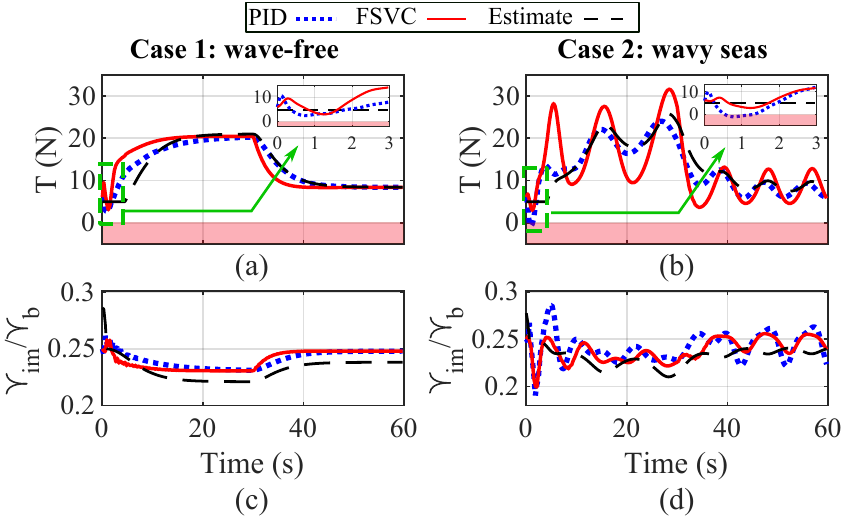}}
\caption{System constraints, $T$ and $\curlyvee_{\text{im}} / \curlyvee_{\mathrm{b}}$, during simulation with the designed FSVC system (red) and a baseline PID controller (blue). The constraint violation region is marked in the red shaded area.} 
\label{fig_T_Vol_im}
\end{figure}
%
On the system constraints side, it is observed in Fig.~\ref{fig_T_Vol_im} that the FSVC system does not violate any, whereas the baseline PID controller violates the \textit{Taut-Cable} constraint, as indicated by the negative cable tension in the $t=[0,3]\SI[unitsep=medium]{}{s}$ window in the top-right corner of Fig.~\ref{fig_T_Vol_im}(b). In the FSVC system, the UAV controls the buoy's forward-surge velocity by directly adjusting the cable tension, $T$; contrarily, the PID controller has no direct control on the cable tension, which explains the constraint violation risk of this controller.
Furthermore, by referring to (\ref{eq_buoy_cable_tension_X}), we see that the FSVC system produces appropriate tension to control the buoy's velocity and minimize the tracking errors presented in Table~\ref{tab_tracking_error_energy}. 

\begin{figure}
\centerline{\includegraphics[width=3.4in]{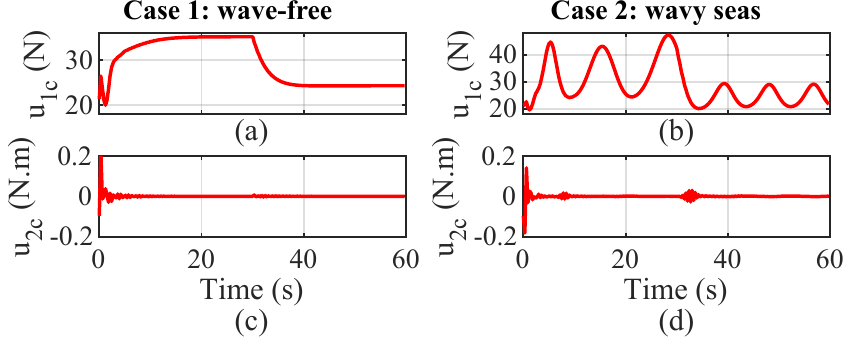}}
\caption{Smooth UAV thrust $u_1$ (top) and torque $u_2$ (bottom) commands generated by the FSVC system in simulation cases $C1$ and $C2$.}
\label{fig_uav_cmd}
\end{figure}

Finally, referring to Fig.~\ref{fig_T_Vol_im}(c) and Fig.~\ref{fig_T_Vol_im}(d), it is observed that the immersed volume variation of the buoy depends on the encounter frequency, which is linked to the `fly over' phenomenon, not to the controller type. Both control systems, FVSC and PID, maintain the buoy's immersed volume between $20-30\%$ in $C1$ and $C2$. In practice, a specific level of immersed volume ratio might be required, such as the application described in \cite{Saleh2019}.
%
%
%
Finally, we note that the FSVC system generates bounded and ringing-free thrust and torque commands to the UAV, as observed in Fig.~\ref{fig_uav_cmd}. This is noteworthy given the relatively complex structure of the FSVC system, yet the PID-like form in (\ref{eq_outer_loop_u_alpha_T}) facilitates tuning the adaptive gains based on well-established PID tuning methods \cite{Kourani2021_PID_Like_Backstepping}.
%
%
%
\par
In summary, the simulations demonstrate that the proposed FSVC system exhibits excellent tracking performance, it attenuates the waves' effect without knowledge of their characteristics, and it offers additive disturbance-rejection to overcome encountered waves, which is highly desirable in marine applications that require precision motion control.
\par
%
\begin{table}
    \caption{Comparison of Tracking Errors and Consumed Energy}
    \begin{center}
    \begin{tabular}{ P{0.7cm} | P{0.8cm} P{0.8cm} | P{0.8cm} P{0.8cm} | P{0.8cm} P{0.8cm}} 
        \hline 
        \multirow{3}{*}{Case} & \multicolumn{2}{c|}{$V$, mean tracking}  & \multicolumn{2}{c|}{$z_{\mathrm{u}}$, mean tracking} & \multicolumn{2}{c}{Total consumed}  \\ 
        & \multicolumn{2}{c|}{error ($\SI[unitsep=medium]{}{cm/s}$)}  & \multicolumn{2}{c|}{error ($\SI[unitsep=medium]{}{cm/s}$)} & \multicolumn{2}{c}{energy ($\SI[unitsep=medium]{}{kJ}$)}  \\ \cline{2-7} 
        &  $PID$   &  $FSVC$ &  $PID$  & $FSVC$  &  $PID$  & $FSVC$ \\
        \Xhline{2\arrayrulewidth}
        $C1$  &  20.0  &  1.3 &  16.1 & 1.8 & 57.2 &  58.1 \\
        $C2$  &  25.6  &  2.0 &  34.6 & 4.8 & 56.1 &  57.0  \\
        \hline
    \end{tabular}
    \label{tab_tracking_error_energy}
    \end{center}
\end{table}
\section{Conclusion} \label{sec_conclusion}
%
The novel problem of a marine locomotive quadrotor UAV was introduced, in which a UAV is tethered to a floating buoy by means of an inextensible cable. 
After deriving the mathematical model of the system components and stipulating
constraints that govern the proposed system's operation, a precision motion control system was designed using polar coordinates to manipulate the forward-surge speed of the buoy, while maintaining the cable in a taut state and keeping the buoy in contact with the water surface. 
The proposed FSVC system was compared in numerical simulations to a baseline Cartesian-based PID controller. The obtained results demonstrated the superior tracking performance and disturbance rejection of the FSVC system in different sea conditions.
\par
The proposed system
paves the way in front of a wide variety of novel marine applications, which can benefit from the superior advantages of multirotor UAVs relative to high speed, maneuverability, ease of deployment, and wide field of vision.
In the future, we aim to extend the problem to the three-dimensional (3D) space, build a representative system prototype, and conduct experimental validation to qualify the obtained simulation results.
\section*{ACKNOWLEDGMENT}

This work is supported by the University Research Board (URB) at the American University of Beirut (AUB).
\bibliographystyle{ieeetr}
\bibliography{Ref/main}
\end{document}